\documentstyle[aaspp4,flushrt]{article}

\def\ergs{erg\,s$^{-1}$}

\def\func1{f(\theta^\prime,\phi^\prime)}
\begin{document}

\title
{
Polarization of SN 1987A Revisited 
}

\author{Lifan Wang\footnote{Also Beijing Astronomical Observatory, Beijing 100080, P. R. China} and J. C. Wheeler}
\affil{Department of Astronomy and McDonald Observatory\\
          The University of Texas at Austin\\
          Austin, TX~78712}

\authoremail{lifan@astro.as.utexas.edu, wheel@astro.as.utexas.edu}

\begin{abstract}
The conventional picture for the origin of the polarization of a supernova
is based on a model of Thomson or resonance scattering of photons 
traveling through an aspherical supernova atmosphere. Positive detection of
intrinsic polarization in SN 1987A is then interpretated as evidence of 
an asymmetrical supernova atmosphere. We show here a different view based 
on the scattering of the supernova light by a dusty circumstellar material 
(CSM), or the ``light echo'' effect. At a given epoch after the 
explosion, the observed photons consist of both those propagating directly 
from the supernova and those scattered by dust particles in the CSM.
Polarized light can be produced if the distribution of the dust
particles is aspherical. The model can reproduce both the time evolution 
of the observed broad band polarization of SN 1987A and major features of 
the polarization spectra. It is also successful
in providing a natural model for the early infrared light curve, in 
particular the observed 30 day delay of the IR maximum compared to the maximum
of the bolometric light curve. \end{abstract}


\section{Introduction}

Polarimetry of SN 1987A was obtained from 
2--262 days after explosion (Jeffery 1991a, and references 
therein). Several groups (Jeffery 1991b; H\"oflich 1991) modelled the
observations in terms of scattering by the supernova atmosphere 
(Sutherland \& Shapiro 1982; McCall 1984). The degree of 
polarization predicted by these models is generally a decreasing 
function of time after explosion, as can be anticipated from the 
fact that the atmosphere gets progressively optically thin as the
supernova atmosphere expands. Eventually  electron scattering becomes 
unimportant and no polarization can be produced. These models are 
capable of reproducing the observations prior to day 100 after 
explosion (H\"oflich 1991; Jeffery 1991b). The observed degree of polarization, 
however,  increased sharply from 0.2\% to about 1.6\% at around day 100 
(Jeffery 1991a; Barret 1988). This behavior is difficult to reconcile
in terms of the photospheric scattering model.

Correction for interstellar polarization (ISP) is crucial, but is unfortunately
quite uncertain. We base our study on the polarimetric data compiled and 
corrected for ISP by Jeffery (1991a), where the ISP is estimated from 
unpublished observations made at La Plata Observatory at around day 600. 
The data show that the degree of polarization increased from 
about 0.1\% at day 2 to about 1\% at $\sim$ day 250. The polarization 
position angle remains fairly constant during all these observations. 
The time variability of the polarization indicates that the observed 
polarization is intrinsically related to the supernova event, regardless of 
the ISP in the direction to the supernova. 

We show here that dust scattering plays an important role in producing the
polarized light of SN 1987A. In \S 2.1 the survival of dust particles in the 
vicinity of the supernova is discussed. The dust scattering 
process and how that leads to the observed polarization is presented in 
\S 2.2. \S 3 compares the model with various observations. Conclusions and 
a brief discussion are given in \S 4.

\section{The Model}

\subsection{Dust Survival}

The dust particles in the 
vicinity of the supernova are destroyed mainly by the intense initial 
ultraviolet (UV) flash of the supernova after the shock breakout. 
Lundqvist \& Fransson (1991), and Luo (1991) inferred that 
an initial UV flash with net energy $\sim\,6\times10^{45}$ ergs and
effective temperatures $\sim\, (3-6)\times10^5$K is required 
to fit the observed light curves of the various nebular emission lines of the 
circumstellar loops. The luminosities and radiation temperatures they used 
were taken from models 10H and 10L of Woosley (1988), and 11E1Y6 of Shigeyama,
Nomoto \& Hashimoto (1988). These models and also those by Arnett (1988) 
give peak luminosities of $(2-3)\,\times\,10^{44}$\ergs and peak 
effective temperatures of $(2-5)\,\times\,10^5$ K. More detailed models by
Ensman \& Burrows (1992) give peak luminosities $(3.7-8)\,\times\,10^{44}$
\ergs and $(5.5-7.2)\,\times\,10^5$ K. The distance at which 
the dust achieves equilibrium temperature $T_d$ is
$$ r_d = {1\over4}\sqrt{ {L\over\pi\sigma T_d^4} {<Q(T_{sn})>\over
<Q(T_d)>} }, \eqno(1)$$
where $T_{sn}$ is the effective temperature of the supernova radiation, 
$<Q(T)>$ is the dust absorption efficiency
averaged over wavelength and is a function of the effective temperature $T$ of
the radiation field, 
and $L$ is the luminosity. Evaporation of dust particles occurs when $T_d$ 
is above the evaporation temperature which is typically around 2000 K. 

As Felten \& Dwek (1989) noticed, however, the evaporation/sublimation 
time scale may be {\it longer} than the duration of
the UV flash, and hence, dust can survive at a substantially smaller
distance from the progenitor than given in equation (1). The time scale 
for dust evaporation at a temperature $T_d$, is (Voit 1992; 
Guhathakurta \& Draine, 1989) 
$$t_{ev} \ \sim\ (3\ \times\ 10^{-11}\ {\rm s})(a/1\ \mu{\rm m}) \exp\ [E_{\rm bind}/kT_d], \eqno(2)
$$
where $E_{\rm bind}$ is the binding energy of the dust particles, and 
$a$ is the radius of the dust particle. Guhathakurta \& Drain (1989)
give $E_{\rm bind}\ \approx 81,000$ K
for graphite and $E_{\rm bind}\ \approx 68,100$ K for silicate. 
At $T_d\ = \ 2000$ K, $t_{\rm ev}$ is $\approx\ 10^{11}$ s, 
considerably larger than the duration of 
the UV flash which is $\le\ 2400$ s. Only grains heated up to a 
temperature at which the corresponding evaporation time is shorter than the 
duration of the UV flash will evaporate. The real temperature above which 
the dust particles will be destroyed is therefore given by equating  
$t_{ev}$ and $t_{UV}$. 

The radius beyond which the dust particles will survive as a function 
of the size of the dust particle is plotted in Fig. 1 for different 
models of the supernova explosions. This defines theoretically the minimum
distance dust particles can survive the initial UV flash. 


\subsection{Dust Scattering}

We define first a polar coordinate system of which the z-axis is along the
line of sight and the x-axis is on the plane of sky and defines the zero point 
of the polarization position angle. The supernova is located at the origin.
The number density of CSM dust is given by $n(r, \theta,\phi)$. 

The scattered light depends sensitively on the shape of the supernova light 
curve. Consider first the case for which the supernova emits natural 
light at a luminosity per unit frequency $L_\nu$ over a time $dt_s$. Assuming 
single scattering and $0\sim dt_s\ << t$, 
the flux of the scattered light $dF_\nu$ is derived by Chevalier (1986).
The Stokes parameters $dQ_\nu$ and $dU_\nu$ can be found in a similar and 
straight forward way. The equations are
$$
\begin{array}{ll}
\left(\begin{array}{c}d F_\nu (t) \\d Q_\nu(t) \\ d U_\nu(t)
  \end{array}\right) \ = \ 
\left(\begin{array}{c} K_1(t) \\ K_2(t) \\ K_3(t)
  \end{array}\right) \ L_\nu dt_s \ =\ & \nl
\int_S dS
\displaystyle{n Q_s \sigma_d\over 4\pi D^2}\ \displaystyle{L_\nu dt_s\over
  4\pi r^2}\left|\displaystyle{dz\over dt}\right|
\left(\begin{array}{c} P_1(\theta) \\ P_2(\theta)\cos 2\phi \\ P_2(\theta)
  \sin 2\phi
 \end{array}\right)
\end{array}
\eqno(3)
$$
where the integration is over the plane of the sky, $Q_s$ is 
the grain scattering efficiency, $\sigma_d$ is the geometric cross
section of the grain, $D$ is the distance to the supernova, and $P_1$ and $P_2$
are elements of the dust scattering matrix (Chandrasekhar 1960) as a function
of $\theta$. The scattering phase function $P_1$ can be found in  
Henvey \& Greenstein (1941), and the polarization function $P_2$ approximates
the Rayleigh function (White 1979).

For the more general case that the supernova light curve is given by
$L_\nu(t_s)$, the fluxes and Stokes parameters can be obtained by convolving 
$L_\nu(t_s)$ with $K_1$, $K_2$ and $K_3$ (cf Chevalier, 1986). 
Scattered light contributes also to the observed luminosity, but is only a  
second order effect in the optically thin case. The net polarization of 
the supernova is therefore calculated by multiplying the polarization of the
scattered light by the ratio of the scattered light to the luminosity of the
supernova.

\section{Comparing with Observations}

Our most favored model is achieved by a dust blob of radius
1.2 $\times\ 10^{16}$ cm located on the plane of the sky passing through 
the supernova, at a distance 5.7 $\times\ 10^{16}$ cm away from 
the supernova and a position angle of $15\arcdeg$. The parameters 
governing dust scattering are: albedo 
$\omega\ =\ 0.5$, and peak linear polarization $P_l\ = \ 0.5$.
The optical depth of the blob is 0.36. In the following this 
model is compared with observations.

\subsection{Broad Band Polarimetry}


We show in Fig. 2(a) the V band polarimetry together with 
the polarimetry curve calculated using equation (3). The V-band light 
curve is from Hamuy \& Suntzeff (1990). The polarimetry data are from Jeffery 
(1991a). The degree of polarization scales linearly
in the adopted optically thin approximation, and satisfactory fits were
obtained for $\tau\ \sim\ 0.2 - 0.5$, depending sensitively on the distance
to the supernova. The dust scattering asymmetry parameter 
is $g$ = 0.3 (Henvey \& Greenstein 1941).
The overall shape of the time evolution of the polarization is well 
reproduced. The increase of polarization before 
day 30 is due to the increase of the flux of the highly polarized 
scattered light as a result of light travel effects. 
The model predicts an increase at around day 100 when the supernova
evolves down from its second optical maximum. This behavior is again due 
to the fact that the fraction of scattered light increases as the supernova 
falls below its maximum. 

\subsection{Polarization Spectra}

The broad spectral features in the polarization spectra which correspond to 
various absorption and emission lines of the supernova were taken as 
evidences againest circumstellar scattering. We show here that this is 
incorrect. The observed polarization depends on both the intrinsic 
supernova light curve and the scattering process in a complicated 
way as shown in equations (3). Because the luminosities at 
different wavelengths evolve differently, the net polarization
at different wavelengths will also be different, especially
across strong P-Cygni lines. It can be anticipated that
broad polarized spectral features can be produced. And indeed, 
as an example, Fig. 2(b) shows the model polarization spectra on day 100 
obtained by using the 
set of spectroscopic observations of SN 1987A from the CTIO archive 
(Phillips et al. 1988). The model polarization spectrum agrees well
with the polarization spectra corrected for 
ISP by Jeffery (1991a), considering the fact that the polarization 
spectra depends sensitively on the corrections for ISP. 

\subsection{Early Infrared Emission}


The dust particles responsible for the scattering and polarization
emit also in the infrared. 
The infrared light curve is calculated by assuming $g\ = \ 0$ 
for the dust scattering phase function. Fig. 3 shows the model results
together with the luminosity of the IR component obtained by fitting
two black bodies to the early photometric observations (Bouchet et al. 
1989) . Although Bouchet et al. (1989) argue that the infrared 
emissions may come from free--free emission of the supernova ejecta, 
the present model correctly fits the light curve of the infrared 
emission, in particular the time delay of its maximum compared with the 
observed light curve. 
Note that the temperature of the dust particles varies with the distance to
the supernova, so their emission can not be exactly that of a black body.
A self-consistent analysis requires detailed modeling of the dust emission
and the early observations which is beyond the scope of this paper.
It is encouraging that such a simple analysis already yields 
surprisingly satisfactory model fits.

\section{Discussions and Conclusions}

The polarimetry of SN 1987A can be modeled 
in terms of scattering by a dusty blob. Both the polarimetry curve and 
the early infrared light curve can be successfully fitted by a CSM 
dust scattering blob.

This analysis is, however, far from complete. Substantial improvements can be
achieved by more detailed modeling. In particular, the 
polarization position angle also shows clear changes across spectral 
lines. This suggests that a single scattering blob can not account 
for all the observed polarization. Two or more blobs may improve the 
model. Another possibility is that the observed polarization is a 
combination of the atmospheric scattering model and the CSM 
dust scattering model presented here. 

The hypothesized dust blob may be related with the mystery spot discovered
in earlier speckle observations (Cropper et al. 1988). They are 
located at almost identical position angles of $\sim\, 15\arcdeg$. Further 
analysis may provide
some interesting tests for the authenticity of the speckle sources 
(Nisenson et al. 1988; Meikle et al. 1987).

Our recent observations show that polarization $\sim\ 1\%$ may be a
common phenomena in Type II supernovae (Wang et al. 1996). A correct picture
for the origin of the polarization is therefore critical for the understanding
of supernova explosion mechanisms in general, and also in 
establishing Type II supernovae as extra--galactic distance indicators
using the expanding photosphere method. If dust scattering is really 
important in producing polarization, the present study suggests that Type II 
supernovae are more likely to be associated with CSM dust than Type Ia.
It is especially likely that dust scattering plays a role in SN 1993J which
was strongly polarized and has a substantial CSM. 
Polarization measurement can thus set strong constraint on the enviroment
of supernovae (Wang \& Wheeler 1996).

This research is supported by NSF Grant AST 9218635 and NASA grant GO 6020
from the Space Telescope Science Institute which is operated by the Association
of Universities for research in Astronomy.

\pagebreak

\noindent
{\bf Figure Captions}\\[1cm]
{\bf Fig. 1} Dust survival radius for different supernova explosion models. The
luminosities (in units of $10^{44}$ erg/sec) and durations (in seconds)
of the out-burst were approximated 
as: 5, 20 (500full1), 2.3, 200 (11E1Y6), and 0.35, 2400 (10H), respectively.
The effective temperature of the out-burst was taken as 5.0$\times 10^{5}$ K 
for all of the models.
The dust absorption coefficients were taken from Draine \& Lee (1984).\\[1cm]
{\bf Fig. 2} (a) V band polarimetry of SN 1987A; (b) Polarization spectrum 
at day 100.\\[1cm]
{\bf Fig. 3} Early infrared observations. The bolometric luminosity was scaled
down by a factor of 10 for clarity.

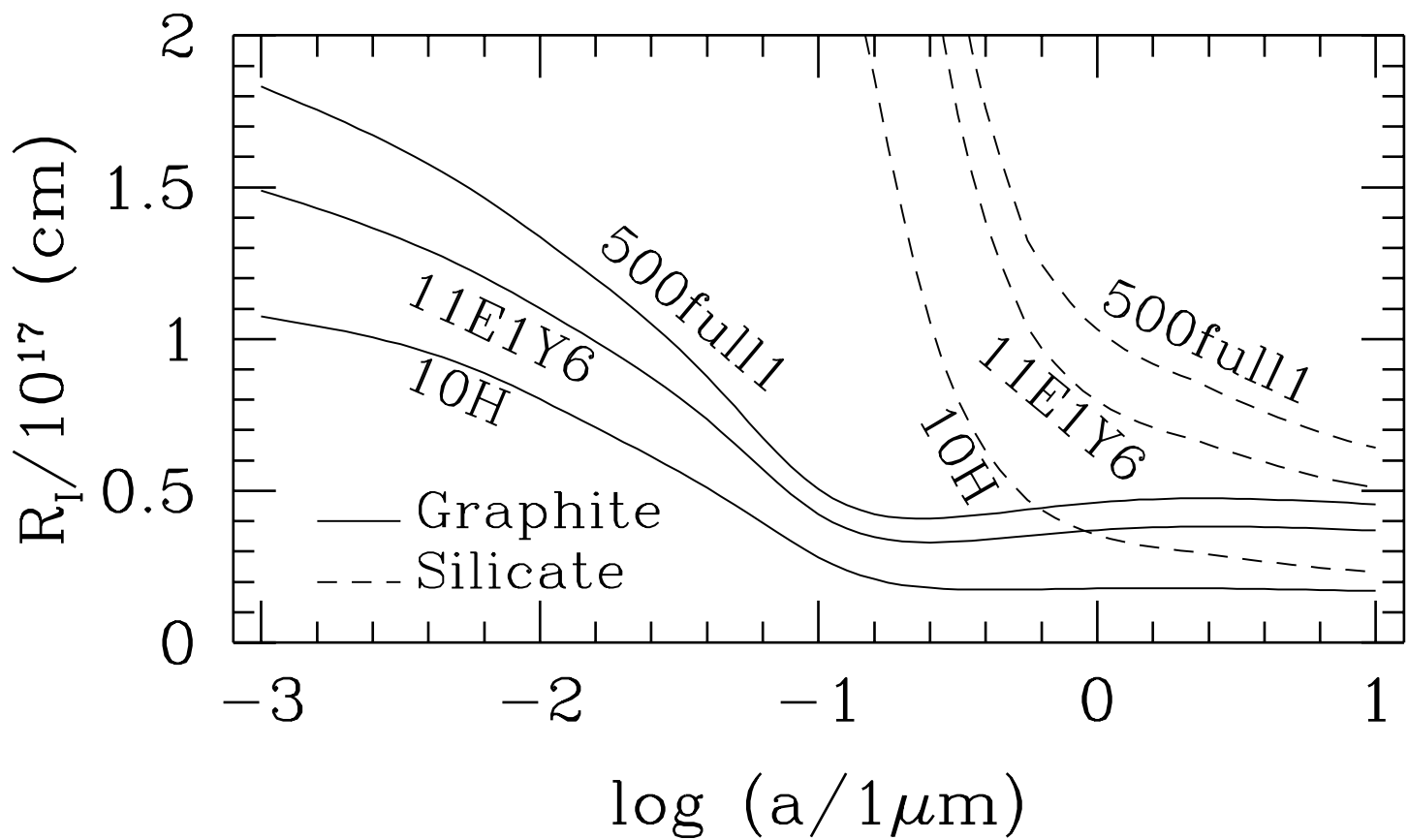
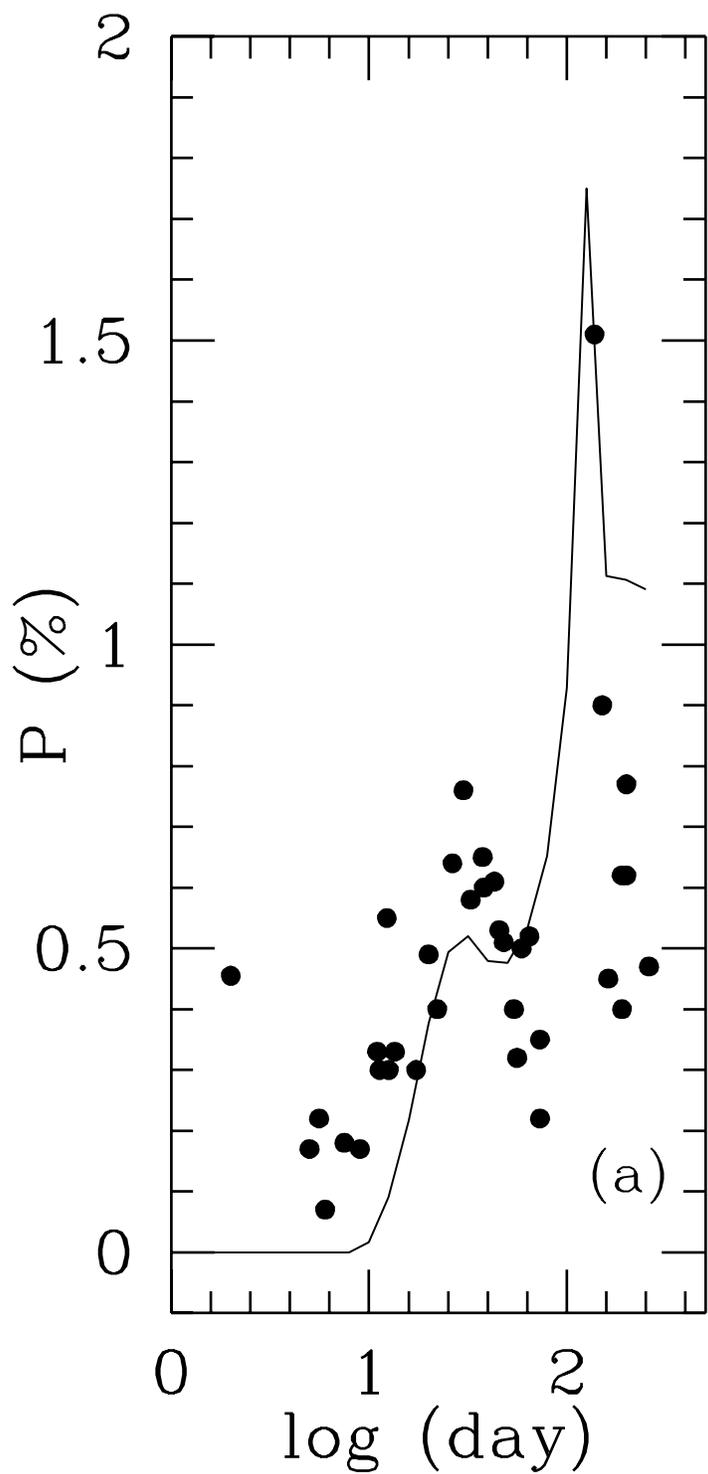
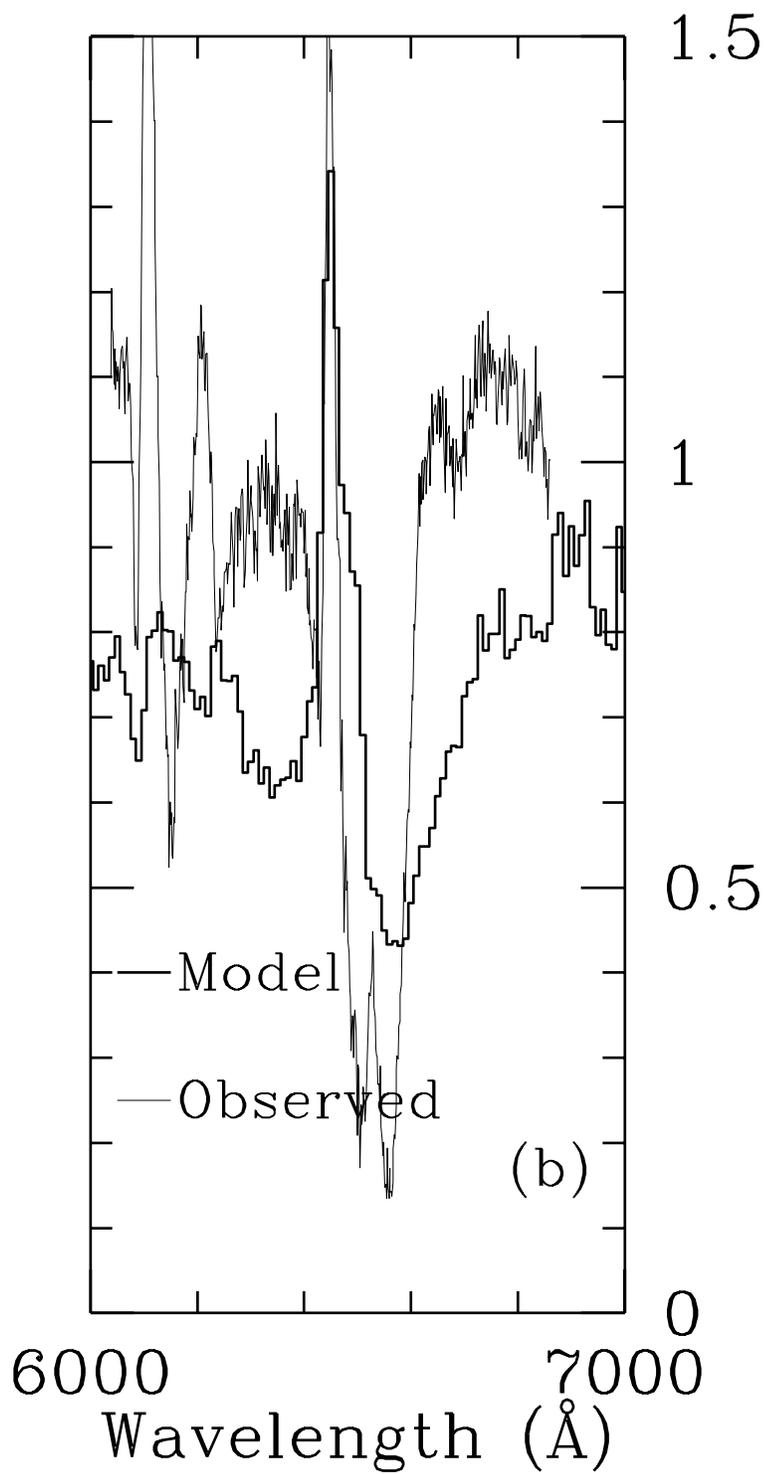
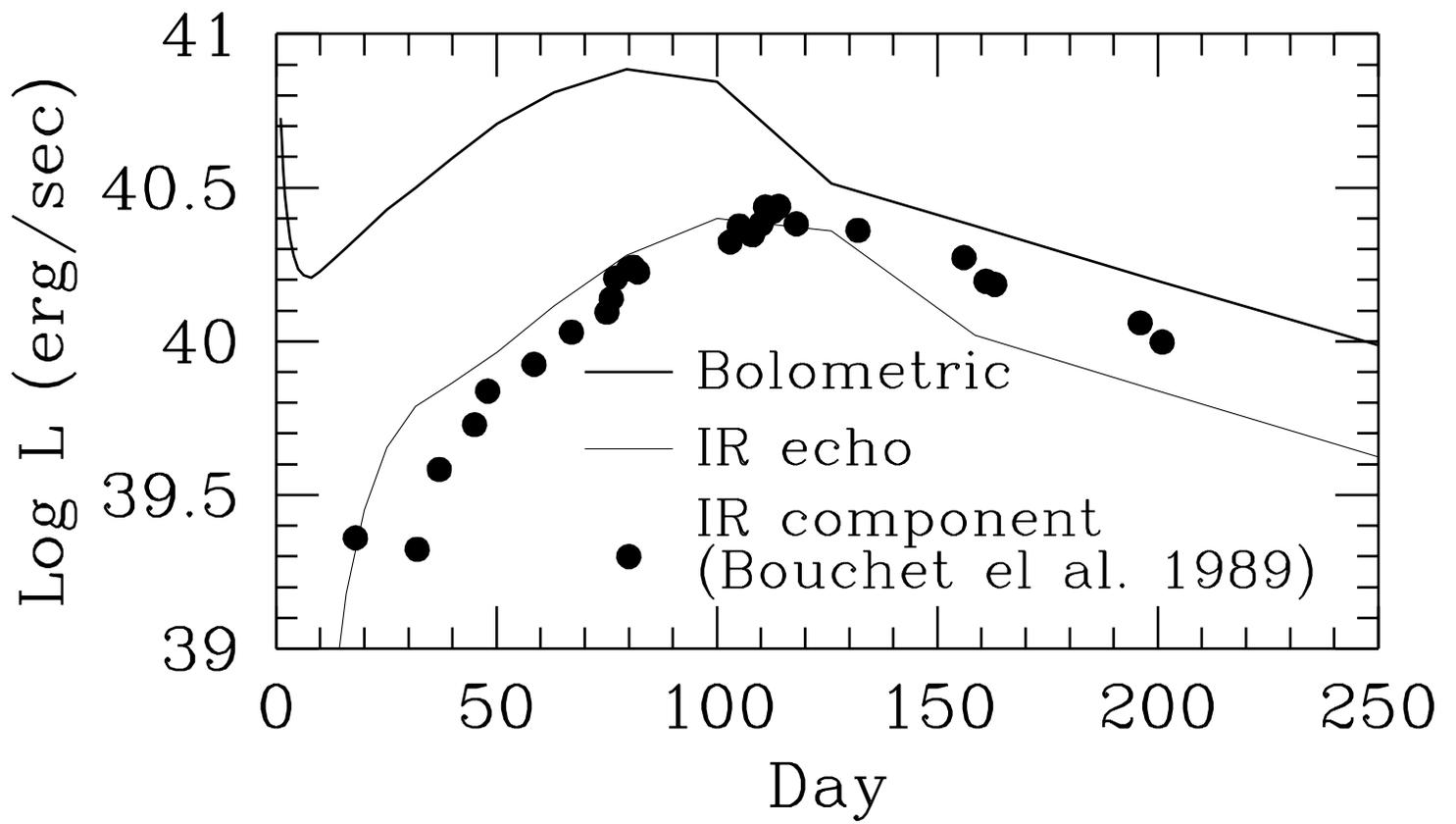
\end{document}